\documentclass[aps,amsmath,prd,preprint,floats,epsf,superscriptaddress,nofootinbib]{revtex4}
\usepackage{graphicx}
\usepackage{bm}
\usepackage{color}
\usepackage{amsmath,amssymb}

\begin{document}

\preprint{IPMU16-0023}
\bigskip

\title{$750$\,GeV diphoton resonance in a visible heavy QCD axion model}

\author{Cheng-Wei Chiang}
\email[e-mail: ]{chengwei@ncu.edu.tw}
\affiliation{Center for Mathematics and Theoretical Physics and Department of Physics, 
National Central University, Taoyuan, Taiwan 32001, R.O.C.}
\affiliation{Institute of Physics, Academia Sinica, Taipei, Taiwan 11529, R.O.C.}
\affiliation{Physics Division, National Center for Theoretical Sciences, Hsinchu, Taiwan 30013, R.O.C.}
\affiliation{Kavli IPMU (WPI), UTIAS, The University of Tokyo, Kashiwa, Chiba 277-8583, Japan}

\author{Hajime Fukuda}
\email[e-mail: ]{hajime.fukuda@ipmu.jp}
\affiliation{Kavli IPMU (WPI), UTIAS, The University of Tokyo, Kashiwa, Chiba 277-8583, Japan}

\author{Masahiro Ibe}
\email[e-mail: ]{ibe@icrr.u-tokyo.ac.jp}
\affiliation{Kavli IPMU (WPI), UTIAS, The University of Tokyo, Kashiwa, Chiba 277-8583, Japan}
\affiliation{ICRR, The University of Tokyo, Kashiwa, Chiba 277-8582, Japan}

\author{Tsutomu T. Yanagida}
\email[e-mail: ]{tsutomu.tyanagida@ipmu.jp}
\affiliation{Kavli IPMU (WPI), UTIAS, The University of Tokyo, Kashiwa, Chiba 277-8583, Japan}

\date{\today}

\begin{abstract}
In this paper, we revisit a visible heavy QCD axion model in light of the recent reports on the $750$\,GeV diphoton 
resonance by the ATLAS and CMS experiments.
In this model, the axion is made heavy with the help of the mirror copied sector of the Standard Model
while the successful Peccei-Quinn mechanism is kept intact.
We identify the $750$\,GeV resonance as the scalar boson associated with spontaneous breaking of the 
Peccei-Quinn symmetry which mainly decays into a pair of the axions.
We find that the mixings between the axion and $\eta$ and $\eta'$ play important roles in its decays
and the resultant branching ratio into two photons.  The axion decay length can be suitable for explaining the diphoton excess by the di-axion production when its decay constant $f_a \simeq 1$\,TeV.
We also find that our model allows multiple sets of the extra fermions without causing the domain wall problem, which is advantageous to explain the diphoton signal.

\end{abstract}


\maketitle

\section{Introduction}
\label{sec:intro}

The success of Kabayashi-Maskawa mechanism\,\cite{Kobayashi:1973fv} for $CP$ 
violation in the quark sector as well as the Sakharov conditions for baryogenesis 
in the early universe\,\cite{Sakharov:1967dj,Fukugita:1986hr} strongly suggest that $CP$ violation is an intrinsic structure of nature. 
If this is indeed the case, it naturally leads to the question why the strong interaction conserves the $CP$ symmetry so well when it is allowed to have its own $CP$-violating parameter, the $\theta$ angle.
This is the strong $CP$ problem in QCD. 

The Peccei-Quinn mechanism\,\cite{Peccei:1977hh,Peccei:1977ur} is the most attractive solution to the strong $CP$ problem.
As a prominent prediction, this mechanism comes with a pseudo-Nambu-Goldstone boson, the axion, 
of mass of $O(100)$ keV\,\cite{Weinberg:1977ma,Wilczek:1977pj}. 
Such a light axion, however, has been excluded by extensive experimental searches\,\cite{Agashe:2014kda}.
  
To circumvent the experimental search constraints, there are two approaches: one way is to make the axion couplings to
the standard-model particles very weak, and the other is to make the axion  heavier. 
The former lead to the well-known invisible axion models\,\cite{Kim:1979if,Shifman:1979if,Zhitnitsky:1980tq,Dine:1981rt} 
in which the axion decay constant $f_a$ is taken to be very large, {\it e.g.}, $f_a>10^9$\,GeV, 
so that the axion couplings are highly suppressed.
The invisible axion models have, however, a serious drawback.
A global $\text U(1)$ symmetry is an essential ingredient of the Peccei-Quinn mechanism. 
However, such a global symmetry is most likely broken explicitly by gravitational effects. 
These effects emerge as higher dimensional operators suppressed by the Planck scale in the effective field theory which shift the $\theta$ angle.
By remembering the stringent upper bound on the $\theta$ angle, $\theta \lesssim 10^{-10}$\,\cite{Agashe:2014kda}, 
the shift in the $\theta$ angle by the gravitational effects is unacceptably large even for the original axion model\,\cite{Weinberg:1977ma,Wilczek:1977pj} where the decay constant is the electroweak scale.
The situation becomes even worse for invisible axion models with $f_a \gtrsim 10^{9}$\,GeV.

In parallel to the invisible axion models, many people have also tried to construct models involving the Peccei-Quinn mechanism with a heavy axion, 
which turns out to be very difficult\,\cite{Tye:1981zy}. 
Among various attempts, however, an exceptionally successful idea was proposed by Rubakov\,\cite{Rubakov:1997vp}
where a mirror world of the Standard Model was introduced.
Recently, a concrete and viable model of such a heavy axion model
has been constructed\,\cite{Fukuda:2015ana} (see also Refs.~\cite{Berezhiani:2000gh,Hook:2014cda,Albaid:2015axa}%
\footnote{The models discussed in  Refs.~\cite{Berezhiani:2000gh,Hook:2014cda} have various unsolved cosmological problems.})
and is called a visible heavy QCD axion model.
In that work, we showed that the axion decay constant could be as low as $f_a\simeq O(1)$\,TeV and axion mass was around 
$m_a>O(0.1)\,{\rm GeV}$ without any conflict with experimental, astrophysical or cosmological constraints.
This model is  visible in the sense that it predicts a scalar boson and vector-like fermions at 
${\cal O}(1)$\,TeV in addition to the heavy axion. 
It should be emphasized that the above-mentioned gravitational breaking effects are sufficiently small, thanks to such a small $f_a$ and the heaviness of the axion. 

In this paper, we revisit a visible heavy QCD axion model in light of the recent reports on the $750$\,GeV diphoton 
resonance by the ATLAS and CMS experiments\,\cite{ATLASdiphoton,CMS:2015dxe}. 
As a result, we show that the scalar boson predicted in the visible axion model with $f_a \simeq 1$\,TeV
can be identified with mass 750 GeV and mainly decays into two axions.
Each axion decays into two photons with a sizable branching ratio, so that the di-axion signal mimics the diphoton signal.
So far, many works have been done to discuss the excess and some of them 
\cite{Higaki:2015jag,Kim:2015xyn,Aparicio:2016iwr,Ellwanger:2016qax} 
(see also ~\cite{Knapen:2015dap,Agrawal:2015dbf,Chala:2015cev,Dasgupta:2016wxw}
for related works) try to explain the signal using axion-like particles. 
Our model is the first realistic model which explains the diphoton excess by using the QCD axion that solves the strong $CP$ problem.

The paper is organized as follows. In section~\ref{sec:visible_axion}, we briefly summarize the status of the heavy axion as well
as the model of Ref.~\cite{Fukuda:2015ana}. 
In section~\ref{sec:interpretation},  we discuss how the model explains the diphoton resonance
by identifying it as the scalar boson in the visible axion model.
We carefully discuss how the axion decays since the scalar boson mainly decays into a pair of the axions.
Finally, discussions and conclusions are given in section~\ref{sec:discussions}.

\section{Heavy QCD Axion model}
\label{sec:visible_axion}

In order to achieve a heavy axion, Rubakov\,\cite{Rubakov:1997vp} proposed an idea to employ a mirror copy of the Standard Model.
By assuming a $\mathbb{Z}_2$ symmetry between the Standard Model and its mirror copy, the $\theta$-angles in these two sectors 
are aligned at the high energy input scale, such as the Planck scale.
The electroweak symmetry breaking scale and the dynamical scale of QCD in the copied sector
can be varied once the $\mathbb{Z}_2$ symmetry is spontaneously broken while the $\theta$ angles are intact.
With the higher scales in the copied sector, the axion mass is enhanced without spoiling the success of the 
Peccei-Quinn mechanism.

In Ref.~\cite{Fukuda:2015ana}, Harigaya and three of the authors propose a concrete realization of this mechanism. 
We prepare the Standard Model with a single Higgs doublet and its mirror copy.
To implement the Peccei-Quinn mechanism, we introduce extra colored vector-like fermions $\psi$ and $\psi'$
in each sector and identify their chiral symmetries as the $\text U(1)$ Peccei-Quinn symmetry. 
Hereafter, objects with a prime (${}'$) refer to those in the copied sector. 
A complex scalar $\phi$ is introduced to break the Peccei-Quinn symmetry spontaneously.  
As in the KSVZ axion model~\cite{Kim:1979if,Shifman:1979if}, $\phi$ couples to  $\psi$ and $\psi'$ via
\begin{eqnarray}
\label{eq:PQ_int}
\Delta \mathcal L = g\phi\left(\bar{\psi}\psi + \bar{\psi}'\psi'\right)\ ,
\end{eqnarray}
where $g$ denotes the coupling constant.
Assuming that the Peccei-Quinn symmetry is spontaneously broken by the vacuum expectation value (VEV) of $\phi$, we decompose $\phi$ into an axion $a$ and a scalar boson $s$ around its VEV, $f_a/\sqrt{2}$,
\begin{eqnarray}
\phi = \frac{1}{\sqrt{2}}(f_a + s) e^{i a/f_a}\ .
\label{eq:phi}
\end{eqnarray}
In terms of the axion, the Peccei-Quinn symmetry is realized non-linearly by
\begin{eqnarray}
a/f_a \to a/f_a +  \alpha \ , \quad \alpha \in [0,2\pi) \ .
\end{eqnarray}

By using a $\mathbb{Z}_2$ breaking spurion $\sigma$, the squared mass parameters of Higgs and Higgs${}'$ 
can be varied with each other~\cite{Fukuda:2015ana}.
Besides, the dynamical scale of QCD ($\Lambda$) and QCD${}'$ ($\Lambda'$) can be also varied 
by introducing some new scalar particles in both sectors whose masses 
are different from each other by the spurion effect.
By making the electroweak$'$ scale and $\Lambda'$ larger than those in the Standard Model sector, 
the axion mass is dominated by the contributions from the copied sector:
\begin{eqnarray}
m_a \propto \frac{f_{\pi'} m_\pi'}{f_a}\ ,
\end{eqnarray}
which can be much larger than conventional models.
Here, $m_\pi'$ and $f_\pi'$ are the mass and the decay constant of the pion$'$ in the copied sector.

In Ref.~\cite{Fukuda:2015ana}, we examined astrophysical and cosmological constraints on the heavy axion model.
We found that the axion heavier than the QCD phase transition temperature,  $T_{\rm QCD} = {\cal O}(100)$\,MeV,
satisfied the astrophysical/cosmological constraints for at least $f_a \lesssim 10^{5}$\,GeV. 
In fact, with such a heavy axion, all the copied sector particles decouple 
from the thermal bath of the Standard Model above the QCD phase transition, 
and hence, the contributions of the copied sector to the effective number of relativistic species are diluted
(see also discussions at the end of section \ref{sec:axion}).%
\footnote{The copied sector is in thermal equilibrium with the Standard Model particles even
after the QCD$'$ phase transition.}

Several remarks on the cosmological constraints are in order.
When we assume the seesaw mechanism\,\cite{Seesaw} in both sectors, the neutrino masses 
in the copied sector are proportional to the Higgs${}'$ VEV squared, leading to too much
(hot/warm) dark matter as the VEV of Higgs${}'$ is about hundreds times larger than the VEV of the Higgs.
To avoid this problem, it is safer to assume that the seesaw mechanism does not take place in the copied sector%
\footnote{This is possible by switching off the $(B-L)'$ symmetry breaking in the copied sector
by using the ${\mathbb Z}_2$ spurion.}, so that the neutrino$'$ decays into pion$'$ and charged lepton$'$ in the copied sector.
The nucleons$'$, electron$'$ and pions$'$ in the copied sector are stable, which places upper limits
on the electroweak$'$ and the QCD$'$ scales in the copied sector to avoid too much dark matter abundance~\cite{Fukuda:2015ana}.
It is also important to arrange the model so that $\psi$ and $\psi'$ mix with the quarks and quarks' 
so that they decay fast enough not to cause cosmological problems.

The constraints from the rare meson decays as well as the beam dump experiments do not
exclude a heavy axion for $m_a \gtrsim 3\times m_{\pi}$.
In this mass range, the axion decay length is  short since it has $a \to 3\pi$ decay modes (see next section),
and hence, it decays well before reaching the detectors of the beam dump experiments such as the CHARM experiment\,\cite{Bergsma:1985qz}.
Besides, due to the lack of direct axion-quark couplings as in the KSVZ type models,
the rate for a rare meson decay into an axion is suppressed by the mixing between the axion and the neutral pion.
As a result, the constraints from the rare meson decays are also insignificant in this mass region.

Let us also consider the constraint on the model from the LHC.
The searches for extra quarks at the LHC put lower limits on the 
mass of an additional fermion, $gf_a/\sqrt{2}$ (see Eq.\,(\ref{eq:PQ_int})).
As alluded to before, $\psi$ decays into quarks via the $\psi$-quark mixing. 
The experimental lower bounds on the masses of extra quarks are 
$800\mbox{--}900\,\text{GeV}$\,\cite{Aad:2014efa,Aad:2015tba,Khachatryan:2015gza,Khachatryan:2015oba}, 
depending on the branching ratios of $\psi$ into $b$ and $t$ quarks.
Assuming $g \sim 1$ if we require it not to blow up below the Planck scale, we find $f_a\gtrsim1\,{\rm TeV}$.

Finally, let us discuss how the Peccei-Quinn mechanism is durable to explicit breaking of the Peccei-Quinn symmetry
by the Planck-suppressed operators.
At the leading order, the explicit breaking terms are  given by dimension-five operators,
\begin{eqnarray}
{\cal L} = \frac{\kappa}{5!M_{\rm PL}} \left(\phi^{5} + \phi^{*5}\right)\ ,
\label{eq:dim5}
\end{eqnarray}
with a coefficient $\kappa$.
Such higher-dimensional operators lead to a non-vanishing $\theta$ angle at the minimum of the 
axion potential:
\begin{eqnarray}
{\mit \Delta} \theta_{\rm eff} \sim 10^{-11} \times\kappa\left(\frac{f_a}{10^{3}\,\rm GeV}\right)^{3}\left(\frac{1\,\rm GeV}{m_a}\right)^2\ ,
\label{eq:shift}
\end{eqnarray}
which is consistent with the current upper limit for $f_a \simeq 1$\,TeV and $m_a =O(1)$\,GeV.
This feature is quite favorable compared with invisible axion models where 
the Peccei-Quinn mechanism can be easily spoiled by explicit breaking terms suppressed by the Planck scale.

\section{Interpretation of the excess}
\label{sec:interpretation}

\subsection{Properties of the scalar resonance}

In this section, we discuss whether the reported $750$\,GeV diphoton resonance can be identified 
with the scalar boson $s$, the radial component of $\phi$ (see Eq.\,(\ref{eq:phi})).
For that purpose, let us first discuss the decay width of $s$.

Assuming that $\psi$ is sufficiently heavy, the effective interactions involving $s$ are given by
\begin{eqnarray}
\label{eq:dilaton_int}
\Delta\mathcal L &=& \frac{s}{f_a}\partial_\mu a\partial^\mu a 
+ \frac{\alpha_s}{8\pi}\frac{g}{\sqrt{2}M_D}f\left(t_D\right) 
sG^{\mu\nu}G_{\mu\nu}
+ \frac{\alpha_2}{8\pi}\frac{g}{\sqrt{2}M_L}f\left(t_L\right) 
sW^{\mu\nu}W_{\mu\nu} 
\nonumber \\
&&+ \frac{2\alpha_Y}{8\pi}
\left(\frac{1}{3}\frac{g}{\sqrt{2}M_D}f\left(t_D\right) 
+\frac{1}{2}\frac{g}{\sqrt{2}M_L}f\left(t_L\right) \right)
sB^{\mu\nu}B_{\mu\nu}  + (G,B,W\to G',B',W')\,,
\end{eqnarray}
where we assume that the copied couplings, $\alpha_s', \alpha_2', \alpha_Y'$, are the same as ones of the Standard Model at UV due to the $\mathbb Z_2$ symmetry and only slightly different at IR.
For a while, we assume that the extra fermions $\psi^{(}{}'{}^{)}$ and $\bar{\psi}^{(}{}'{}^{)}$ form respectively the 
${\mathbf 5}$ and ${\mathbf 5}^*$ representations of $SU(5)_{\rm GUT}$, and name the colored fermion
and doublet fermion, ${\psi^{(}{}'{}^{)}}_{D,L}$, respectively.
The parameters with the subscripts $D,L$ are for ${\psi^{(}{}'{}^{)}}_{D,L}$, respectively.
The three gauge field strengths in each sector are denoted by $G^{(}{}'^{)}$,
$W^{(}{}'^{)}$, and $B^{(}{}'^{)}$, respectively.
The function $f(t)$ is defined by%
\footnote{In the limit of heavy fermion masses, $f(t)$ converges: $\displaystyle{\lim_{t\to\infty}}f(t) \to 2/3$.}
\begin{eqnarray}
\label{eq:loop_func}
f(t)\equiv t\left[1 + (1 - t)\arcsin^2\left(\frac1{\sqrt t}\right)\right]\ , 
\end{eqnarray}
with $t$'s being
\begin{eqnarray}
t_{D,L} = \frac{4M_{D,L}^2}{M_s^2}\ .
\end{eqnarray}
The mass of $s$, $M_s$, is taken to be $750$\,GeV.

As  is clear from this effective Lagrangian, we find that $s$ mainly decays into a pair of the axions,
while the modes into gauge bosons are loop-suppressed.
As a result, the total decay width of $s$ is roughly given by
\begin{eqnarray}
\label{eq:dilaton_width}
\Gamma(s\to 2 a) \simeq \frac{{M_s}^3}{32\pi{f_a}^2} \simeq 4.2\,\text{GeV}\times
\left(
\frac{1\,\rm TeV}{f_a}
\right)^2
\end{eqnarray}
for $M_s = 750\,\text{GeV}$.
It should be noted that it is difficult to explain the broad width $\Gamma\simeq45\,\text{GeV}$ that is slightly favored by the ATLAS experiment unless the decay constant $f_a \lesssim 300$\,GeV.
With such a small decay constant, however, the extra colored particles are predicted to be too light.

At a first glance, it seems that this dilaton $s$ cannot account for the diphoton signal 
since the $2\gamma$ decay mode of $s$ is highly suppressed, $\text{BR}(s\to\gamma\gamma)\sim(\alpha/4\pi)^2\sim10^{-8}$. 
However, the $2a$ decay mode can mimic the diphoton signal since the axions in the final state 
are highly boosted and decay into collimated photons. 
Such photon jet events have been studied in Refs.~\cite{Draper:2012xt,Ellis:2012zp,Ellis:2012sd,Dobrescu:2000jt,Toro:2012sv,Chang:2006bw,Curtin:2013fra}, and for recent works~\cite{Knapen:2015dap,Agrawal:2015dbf,Chala:2015cev,Aparicio:2016iwr,Ellwanger:2016qax,Dasgupta:2016wxw} 
discuss this possibility in the context of $750\,\text{GeV}$ diphoton excess.

At the LHC, the resonance $s$ is produced via the gluon fusion process.
By using the narrow width approximation,  the production cross section of $s$ is estimated to be
\begin{eqnarray}
\sigma(g+g\to s) \simeq 11\,{\rm fb} \times \left(\frac{f(t_D)}{2/3}\right)^2 \left(\frac{1\,\rm TeV}{f_a}\right)^2\ ,
\end{eqnarray}
where we use the {\tt MSTW2008} parton distribution functions~\cite{Martin:2009iq}.%
\footnote{See, {\it e.g.}, Ref.~\cite{Baglio:2010ae} for a discussion on higher-order QCD corrections. }

Therefore, to account for the observed excess 
\begin{eqnarray}
4\,{\rm fb} \lesssim \sigma\times BR(s\to 2a)\times BR(a\to 2\gamma)^2 
\lesssim 10 \,{\rm fb}\ ,
\label{eq:signal}
\end{eqnarray}
the axion branching ratio into photons should be of ${\cal O}(1)$.
As will be shown below, $BR(a\to 2\gamma) = {\cal O}(1)$ is achieved through the mixings between
the axion and the $\eta$ and $\eta'$ mesons in the Standard Model.%
\footnote{To avoid confusion, we reserve the name $\eta'$ to denotes the pseudoscalar meson in the 
Standard Model, and we use $(\eta)'$ and $(\eta')'$ to denote the pseudoscalar mesons
in the copied sector.}

\subsection{Properties of the axion}
\label{sec:axion}

To explain the diphoton excess by mimicking a single photon with collimated photons from the boosted axion decay, the axion needs to have $BR(a\to2\gamma) = {\cal O}(1)$ .
For a very heavy axion higher than the QCD scale, $m_a \gg \Lambda$, the effective couplings of the axion to the Standard Model particles are given by
\begin{eqnarray}
{\cal L}_a =
 \frac{\alpha_s}{8\pi}\frac{a}{f_a} G\tilde{G}+
 \frac{\alpha_2}{8\pi}\frac{a}{f_a} W\tilde{W}
+ 2\left(\frac{1}{3} + \frac{1}{2}\right) 
\frac{\alpha_Y}{8\pi}\frac{a}{f_a} B\tilde{B}\ ,
\label{eq:anomH}
\end{eqnarray}
where we take the limit of $M_{D,L} \gg m_a$.
In this case, the main decay mode of the axion is QCD jets, and the branching ratio into $2\gamma$
is highly suppressed:
\begin{eqnarray}
BR(a\to 2\gamma) \propto \left({a_Y}/{a_s}\right)^2 = {\cal O}(10^{-2})\ .
\end{eqnarray}
Thus, in order to have a sizable branching ratio into the photon mode, the axion mass 
is required to be within ${\cal O}(1)$\,GeV so that the hadronic decay modes are suppressed.
In the following, we examine the branching ratios of the axion for $m_a \lesssim 1$\,GeV,
where the axion mixings to the $\eta$ and $\eta'$ in the Standard Model play important roles.

To discuss how the axion mixes with $\eta$ and $\eta'$ in the Standard Model sector, 
let us consider the effective Lagrangian 
\begin{eqnarray}
{\cal L} &=& \frac{1}{2} \partial_\mu a \partial^\mu a 
+\frac{1}{2} \partial_\mu \eta_8 \partial^\mu \eta_8 
+\frac{1}{2} \partial_\mu \eta_0 \partial^\mu \eta_0 \nonumber\\
&&-\frac{1}{2}m_a^2\, a^2 
- \frac{1}{2}m_8^2\, \eta_8^2 - {\mit \Delta}^2 \eta_8\eta_0
 - \frac{1}{2}m_0^2\left(\eta_0 + \frac{f_0}{\sqrt{6}f_a} a\right)^2 \ ,
\label{eq:eff0}
\end{eqnarray} 
where $\eta_8$ and $\eta_0$ denote the neutral component of the octet 
and the singlet pseudo-Nambu-Goldstone modes resulting from chiral symmetry
breaking of QCD, respectively.
We neglect the contributions from $\pi^0$ because the mixing to $\pi^0$ is more suppressed.
The parameter $f_0$ denotes the decay constant of $\eta_0$, and will be fixed to 
$f_0 \simeq f_\pi \simeq 93$\,MeV shortly.
At the leading order of the chiral perturbation theory, the mass parameters $m_8^2$ and ${\mit \Delta}^2$
are generated from the quark mass terms (see, {\it e.g.}, Ref.~\cite{Beisert:2003zs})
\begin{eqnarray}
m_8^2 &\simeq & \frac{1}{3}\frac{(m_u + m_d + 4m_s)}{(m_u+m_d)} m_\pi^2\ , \\
{\mit\Delta}^2 &\simeq & \frac{\sqrt{2}}{3}\frac{(m_u + m_d -2m_s)}{(m_u+m_d)} m_\pi^2\ ,
\label{eq:CHPT}
\end{eqnarray}
while $m_0^2$ is generated by the anomalous breaking of $U(1)_A$.%
\footnote{Here, we neglect the mass term of $\eta_0$ generated from the quark mass terms.}

The mass term proportional to $m_0^2$ reflects the fact that the combination of the chiral rotation and 
the Peccei-Quinn symmetry,
\begin{eqnarray}
\frac{\eta_0}{f_0} \to \frac{\eta_0}{f_0} + \alpha\ , \quad \frac{a}{f_a} \to \frac{a}{f_a} -\sqrt{6} \alpha\ , \quad (\alpha \in [0,2\pi) )\ ,
\end{eqnarray}
is free from the anomaly of QCD.%
\footnote{We take the normalization of $\eta_0$ such that the $U(1)_A$ symmetry corresponds to
\begin{eqnarray}
\frac{\eta_0}{f_0} \to \frac{\eta_0}{f_0} + \alpha\,\quad &\longleftrightarrow& \quad q_{L,R} \to (e^{i\alpha/\sqrt{6}},\, e^{i\alpha/\sqrt{6}},\,e^{i\alpha/\sqrt{6}})\times q_{L,R}  \ ,
\end{eqnarray}
where $q_{L,R} = (u_{L,R}, d_{L,R}, s_{L,R})$.
}
It should be also noted that in the KSVZ models, the axion appears only through the $m_0^2$ term, 
and there is no kinetic mixing between the axion and the pseudo-Nambu-Goldstone modes.

Since the chiral symmetries and the Peccei-Quinn symmetry have the anomalies of QED, $\eta_{0,8}$ and the axion have anomalous couplings to the photons: 
\begin{eqnarray}
{\cal L} =  \frac{\eta_8}{f_8} \frac{\alpha_{\rm QED}}{4\sqrt{3}\pi}F\tilde{F} +
\frac{\eta_0}{f_0} \frac{\alpha_{\rm QED}}{\sqrt{6}\pi}F\tilde{F} 
+ \frac{8}{3}\frac{a}{f_a} \frac{\alpha_{\rm QED}}{8\pi}F\tilde{F}   \ ,
\label{eq:anoms}
\end{eqnarray}
where $f_8$ denotes the decay constant of $\eta_8$.
The anomalous coupling of the axion comes from the anomalous couplings in Eq.\,(\ref{eq:anomH}).

Now, let us first resolve the mass mixing between $\eta_8$ and $\eta_0$ to obtain the mass eigenstates, $\eta$  and $\eta'$.
Unfortunately, it is known that the observed decay widths of $\eta$ and $\eta'$ into two photons cannot 
be reproduced by using the mass mixing parameter ${\mit\Delta}^2$ in Eq.\,(\ref{eq:CHPT}).%
\footnote{From the calculation of the chiral perturbation theory, $f_8$ is fixed 
to be about $1.3\times f_\pi$ \cite{Donoghue:1986wv,Bijnens:1988kx}.
}
Thus, we follow instead the phenomenological approaches taken in Refs.~\cite{Donoghue:1986wv,Bijnens:1988kx}, 
in which the mixing angle between 
$\eta$--$\eta'$  and $f_0$ are taken to reproduce the observed values using the anomalous couplings
in Eq.\,(\ref{eq:anoms}).
According to Refs.~\cite{Donoghue:1986wv,Bijnens:1988kx}, we find 
\begin{eqnarray}
\sin\theta \simeq -1/3\ , \quad f_0 \simeq  1\times f_\pi \ ,
\end{eqnarray}
where the mixing angle $\theta$ is defined by 
\begin{eqnarray}
\eta_8 &=& \cos\theta\, \eta + \sin\theta\, \eta' \ , \\
\eta_0 &=& -\sin\theta\, \eta + \cos\theta\, \eta' \ .
\label{eq:etaeta}
\end{eqnarray}

Next, let us resolve the axion mixing with $\eta$ and $\eta'$.
From the mass term in Eq.\,(\ref{eq:eff0}), we  find the mass eigenstates $(\eta_D,\eta'_D,a_D)$;
\begin{eqnarray}
\eta &\simeq& \eta_D + \varepsilon_{a\eta} a_D \ ,\\
\eta' &\simeq& \eta'_D + \varepsilon_{a\eta'} a_D \ , \\
a &\simeq& a_D -  \varepsilon_{a\eta} \eta_D -  \varepsilon_{a\eta'} \eta'_D \ ,
\end{eqnarray}
where 
\begin{eqnarray}
\varepsilon_{a\eta} = 
-{}\frac{f_0}{\sqrt{6}f_a} \frac{m_{\eta'}^2}{m_{\eta}^2-m_a^2}  \sin\theta\ ,\quad
\varepsilon_{a\eta'} = 
{}\frac{f_0}{\sqrt{6}f_a} \frac{m_{\eta'}^2}{m_{\eta'}^2-m_a^2}  \cos\theta\ .
\end{eqnarray}
Due to the above mixing, a coupling constant, $c_{a{\cal O}}$,  of $a_D$ to an operator ${\cal O}$
is 
given by,
\begin{eqnarray}
c_{a\cal O}  = {\varepsilon}_{a\eta'} c_{\eta'\cal O} +  {\varepsilon}_{a\eta} c_{\eta\cal O} \ .
\label{eq:combine}
\end{eqnarray}
For example, the axion decay width receives important contributions from the anomalous couplings
of $\eta$ and $\eta'$.

\begin{figure}[tbp]
	\centering
	\includegraphics[width=8cm]{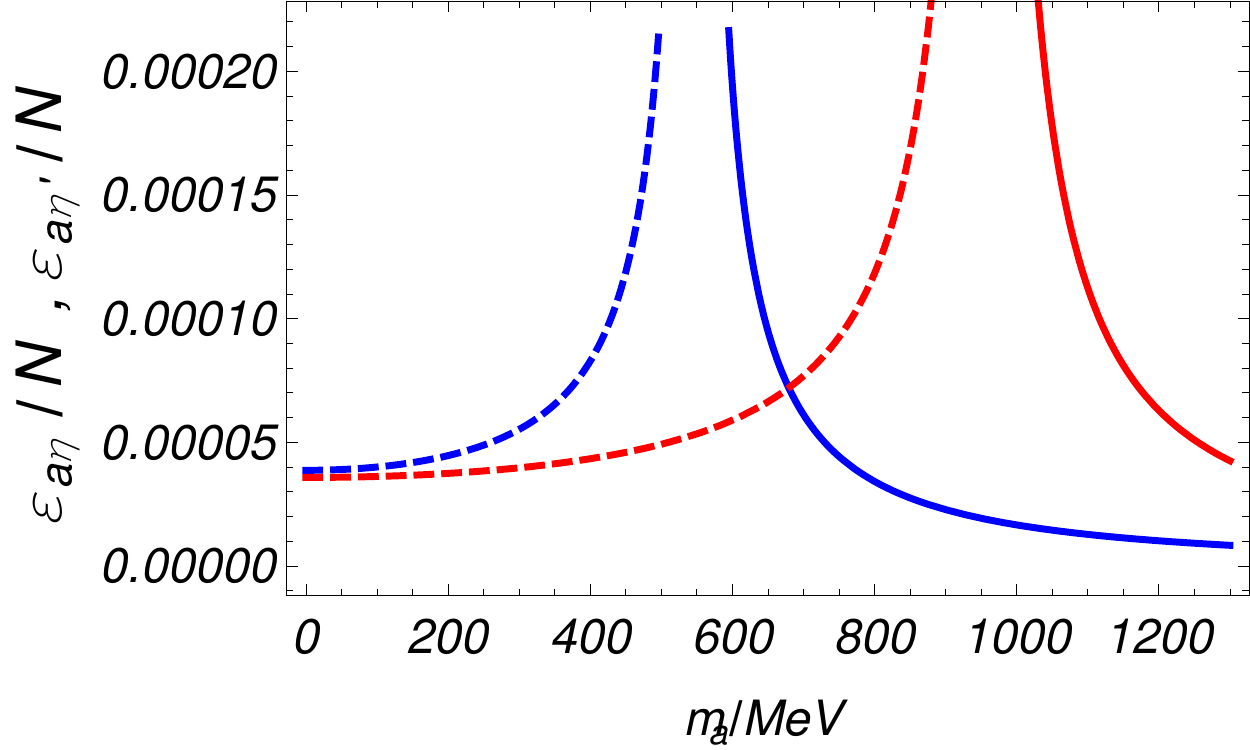}
	\caption{The mixing angles $\varepsilon_{a\eta}$ (blue) and $\varepsilon_{a\eta'}$ (red)
	as a function of $m_a$ for $f_a = 1$\,TeV. 
	The dashed curves indicate the negative mixing angles.	
	Here, we take $\sin\theta \simeq -1/3$ and $f_0 \simeq f_\pi$. Here and hereafter, $N$ denotes the number of colored Fermions, which we will discuss later.
	}
	\label{fig:mixing}
\end{figure}
In Fig.\,{\ref{fig:mixing}}, we show the mixing angles of the axion to $\eta$ and $\eta'$
as a function of $m_a$ for $f_a = 1$\,TeV. 
Here, we take $\sin\theta \simeq -1/3$ and $f_0 \simeq f_\pi$.
The figure shows that the mixing angles are enhanced for either $m_a \simeq m_\eta$ 
or $m_a \simeq m_{\eta'}$.
The signs of the mixing angles are opposite in the region of $m_{\eta} < m_a < m_{\eta'}$.

\begin{figure}[t]
\begin{center}
\begin{minipage}{.46\linewidth}
  \includegraphics[width=\linewidth]{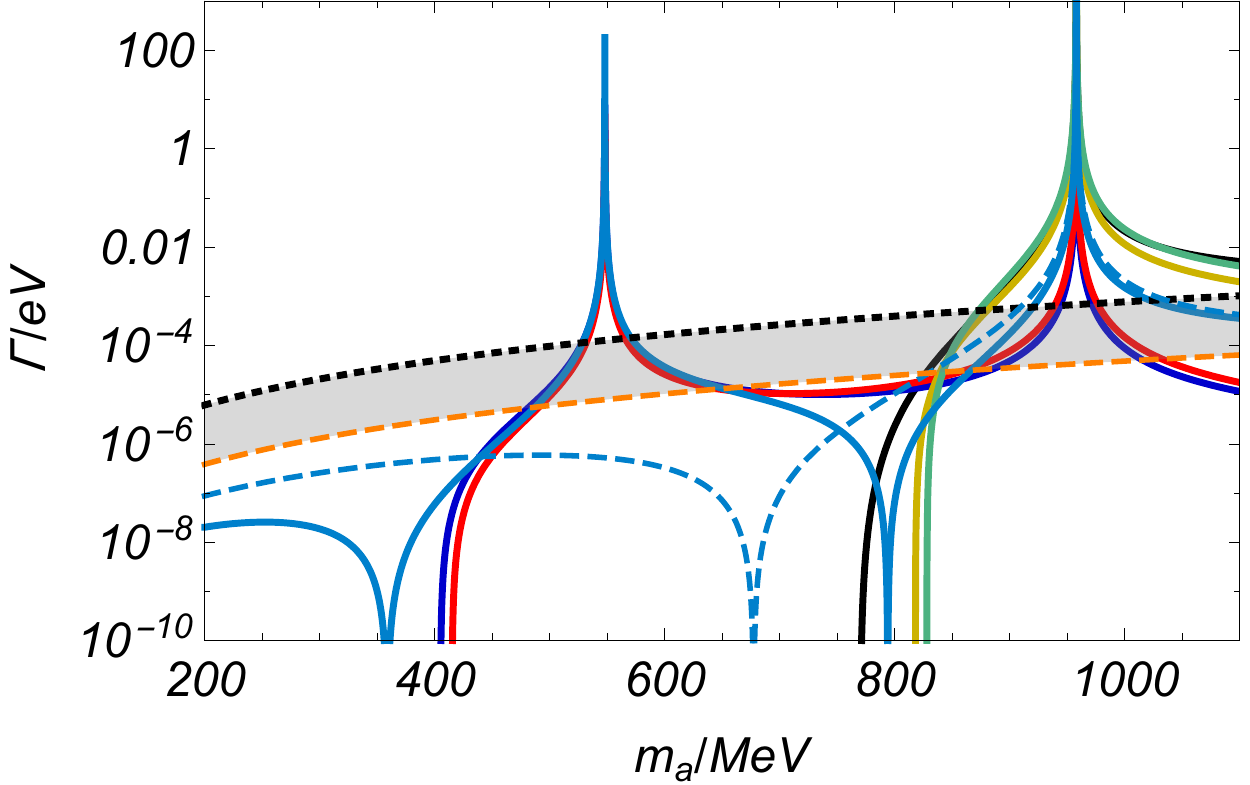}
 \end{minipage}
 \hspace{1cm}
 \begin{minipage}{.46\linewidth}
  \includegraphics[width=\linewidth]{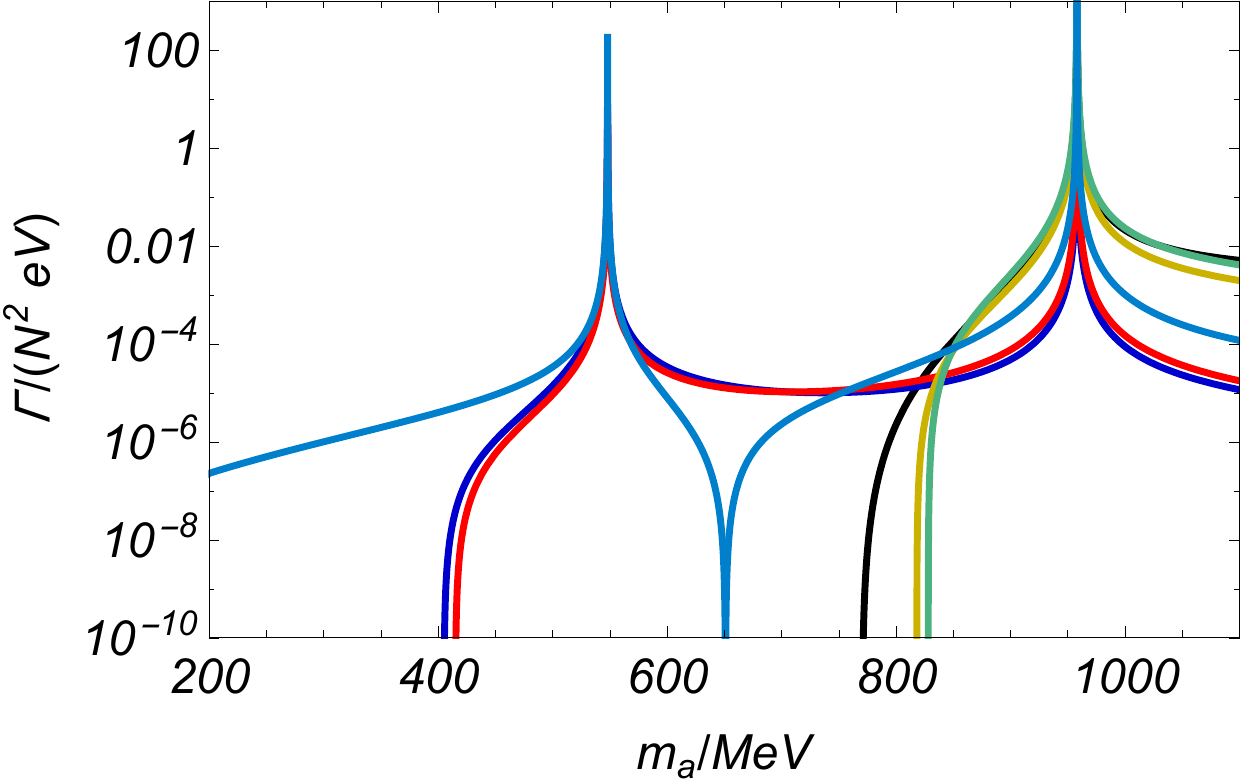}
 \end{minipage}
 \end{center}
\caption{\sl \small
(Left) The relevant partial decay widths of the axion into $\gamma\gamma$ (light blue), 
$3\pi^0$ (dark blue), $\pi^0\pi^+\pi^-$ (red), 
$\rho+\gamma$ (black), $\eta +2\pi^0$ (grass green),
$\eta + \pi^+ + \pi^-$ (green),
and $\gamma' \gamma'$ (black dashed and gray band) for $f_a = 1$\,TeV.
Here we take the $\eta$--$\eta'$ mixing angle to have $\sin\theta \simeq -1/3 $.
The blue dashed line indicates $\Gamma(a\to\gamma\gamma)$ without the $a\mbox-\eta$ mixing. Here we can see that the dip around $650\,\text{MeV}$ comes from the phase cancellation.
The orange dashed curve represents the decay width of the 
axion into $\gamma\gamma$ only from the intrinsic axion anomalous coupling to $\gamma$ in Eq.\,(\ref{eq:anoms}). In this figure, we take $N=1$.
(Right) The partial decay widths for vanishing intrinsic axion anomalous couplings to $\gamma$ and $\gamma'$.
}
\label{fig:width}
\end{figure}

By taking into account the mixing effects, we calculate the decay widths of the relevant modes, as shown in Fig.\,\ref{fig:width}.
Two figures show that the widths are enhanced for either $m_a \simeq m_\eta$ or $m_a \simeq m_{\eta'}$, where the mixing to $\eta$ or $\eta'$ is enhanced. 
The figure also shows that the $2\gamma$ mode is suppressed at around $m_a \simeq 670$\,MeV,
which is due to destructive interference between the contributions of $\eta$ and $\eta'$. 
The analysis of each mode is given in the appendix.

Several comments are in order.
As discussed in the appendix, we estimate the $3\pi$ and $\eta+2\pi$ modes from 
the observed decay widths of $\eta$/$\eta'$ into 3$\pi$ and $\eta'$ into $\eta+2\pi$.%
\footnote{It should be noted that it is difficult to reproduce the observed decay widths
using the chiral perturbation theory (see, for example, Ref.~\cite{Ecker:2015uoa}).}
We approximate the decay amplitudes squared by the decay widths divided by the phase space volume.
We then combine them according to Eq.\,(\ref{eq:combine}), assuming that they add up constructively.
Accordingly, our estimates of the widths into $3\pi$ and $\eta+2\pi$ have ${\cal O}(1)$ uncertainties.%
\footnote{In fact, the observed Dalitz plots of $\eta\to\pi^0\pi^+\pi^-$ and $\eta'\to3\pi^0$ are not flat 
\cite{Ablikim:2015cmz}.}

In the figure, we also show the decay width of the $2\gamma'$ mode as a gray band.
Similar to the width of the $2\gamma$ mode, this mode also receives contributions 
from the mixing of the axion to $(\eta)'$, $(\eta')'$ and $CP$-odd glueball$'$ in the copied sector,
depending on the mass parameters.
In this paper, we parametrize the anomalous coupling to the $\gamma'$ by
\begin{eqnarray}
{\cal L} \simeq  \left(\frac{8}{3} + A_{\rm eff}\right)\frac{a}{f_a} \frac{\alpha_{\rm QED'}}{8\pi}F'\tilde{F}'   \ ,
\label{eq:aF'F'}
\end{eqnarray}
where $\alpha_{\rm QED'} \simeq \alpha_{\rm QED}$ by assumption.
It should be noted that ${A}_{\rm eff}$ is highly suppressed when $m_{u',d'} \gg \Lambda'$
and  the mass of the axion mainly comes from the mixing to the $CP$-odd glueball$'$ in the copied sector.
The maximal value of $A_{\rm eff} \simeq 8$ which is the case when the axion mainly mixes with $(\eta_0')'$
and there is no mixing between $(\pi^0)'$ and $(\eta_8)'$.

\begin{figure}[t]
\begin{center}
  \includegraphics[width=.55\linewidth]{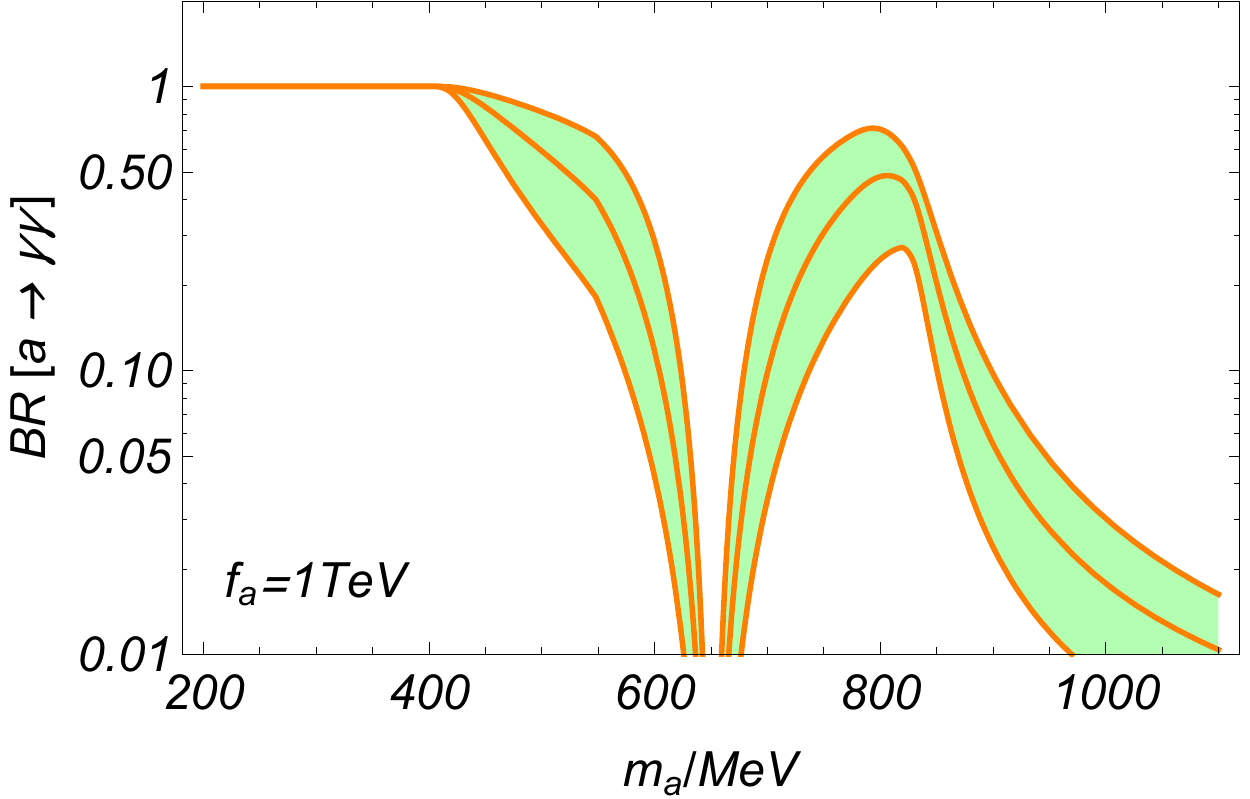}
 \end{center}
\caption{\sl \small
The branching ratio of the axion into $\gamma\gamma$ for $f_a = 1$\,TeV.
We set the $\eta$--$\eta'$ mixing angle to be $\sin\theta \simeq -1/3 $.
The central curve in the band corresponds to $BR(a\to\gamma\gamma)$
when we estimate the widths into $3\pi$ and $\eta+2\pi$ by using 
the approximation given in the appendix, while the upper and the lower curves
correspond to the width multiplied by a factor of $1/3$ and $3$, respectively.
We assume that the $a\to\gamma'\gamma'$ width is zero due to the anomaly cancellation.
}
\label{fig:branching}
\end{figure}

As we discussed in \cite{Fukuda:2015ana},  $\gamma'$ is required to decouple from
the thermal bath of the Standard Model before the hadron decoupling, which requires
that the axion mass is much higher than the QCD phase transition temperature.
Since the mass range of our interest is not very much larger than the QCD phase transition scale,
the decoupling of  $\gamma'$ could have not completed before the QCD phase transition,
since the Boltzmann suppression of the axion production is not significant in this mass range. 
In fact, for $m_a \lesssim 1$\,GeV, we find that the decoupling of  $\gamma'$ completes at a
temperature slightly below the QCD phase transition temperature when the axion anomalous coupling to the $\gamma'$
is given by the one in Eq.\,(\ref{eq:aF'F'}).
To evade the dark radiation constraint, $N_{\rm eff} = 3.15 \pm 0.23$ \cite{Ade:2015xua}, we hereafter 
assume $A_{\rm eff} \ll 1$ and also introduce extra matter fields in 
both sectors which couple to $\phi$ so that the intrinsic anomalous coupling is canceled. Hence, $N$, the total number of colored Fermions is need not to be unity. Then, the axion-gluon couplings in Eqs.\,(\ref{eq:eff0}) and (\ref{eq:anoms})
as well as the gluon couplings of $s$ in Eq.\,(\ref{eq:dilaton_int}) are multiplied by $N$.
We will mention the domain wall problem associated to non-unity $N$ later.
Accordingly, the intrinsic axion anomalous coupling to $\gamma$ ({\it i.e.}, the last term in Eq.\,(\ref{eq:anoms})) is 
also vanishing.
In the right panel of Fig.\,\ref{fig:width},  we show the branching ratios assuming vanishing intrinsic 
axion anomalous couplings to both $\gamma$ and $\gamma'$.
We also take $A_{\rm eff} \ll 1$ assuming $m_{u',d'} \gg \Lambda'$ in the copied sector.
The figure shows that vanishing anomalous coupling case is rather preferred in the sense of enough branching ratio.

\subsection{$750$\,GeV di-axion resonance}

As discussed above, the axion has sizable decay widths into $2\gamma$ and $3\pi^0$.
When the axion is highly boosted, these modes seem to mimic the diphoton signal
since each $\pi^0$ immediately decays into $2\gamma$.
However, the boosted $3\pi^0$ is not acceptable as a single photon.  This is
because when the number of photons is large, some of them are converted before the electromagnetic calorimeter.  And once their energies are measured, it is difficult for them to fake a single photon signal\footnote{
For a single photon, the cumulative conversion rate before it
reaches to the electromagnetic calorimeter is about 20\,\% 
even in the central region of the ATLAS detector ({\it i.e.}, $\eta = 0$)~\cite{ATLASNOTE}.
Accordingly, the conversion probability of a photon-jet consisting of six photons 
is naively expected to be 100\,\% for a promptly decaying axion, where the 
charged tracks made by the conversion carry only a fractional energy of the axion.
For a detailed analysis on the discrimination of photon-jets 
from single photons at the LHC detectors, see \cite{withMihoko}.}.
It should be also noted that it is not obvious even whether a collimated $2\gamma$ event is acceptable as a single photon.
Considering the detector inner structure\,\cite{Aad:2008zzm,Chatrchyan:2008aa}, we conclude that the $2\gamma$ mode is safe and does not reduce the acceptance of the signal 
because their energies are high and the converted $e^+e^-$ are not bent too much for the tracker to measure the energy.  A more detailed study will be given elsewhere.

\begin{figure}[t]
\begin{center}
  \includegraphics[width=0.55\linewidth]{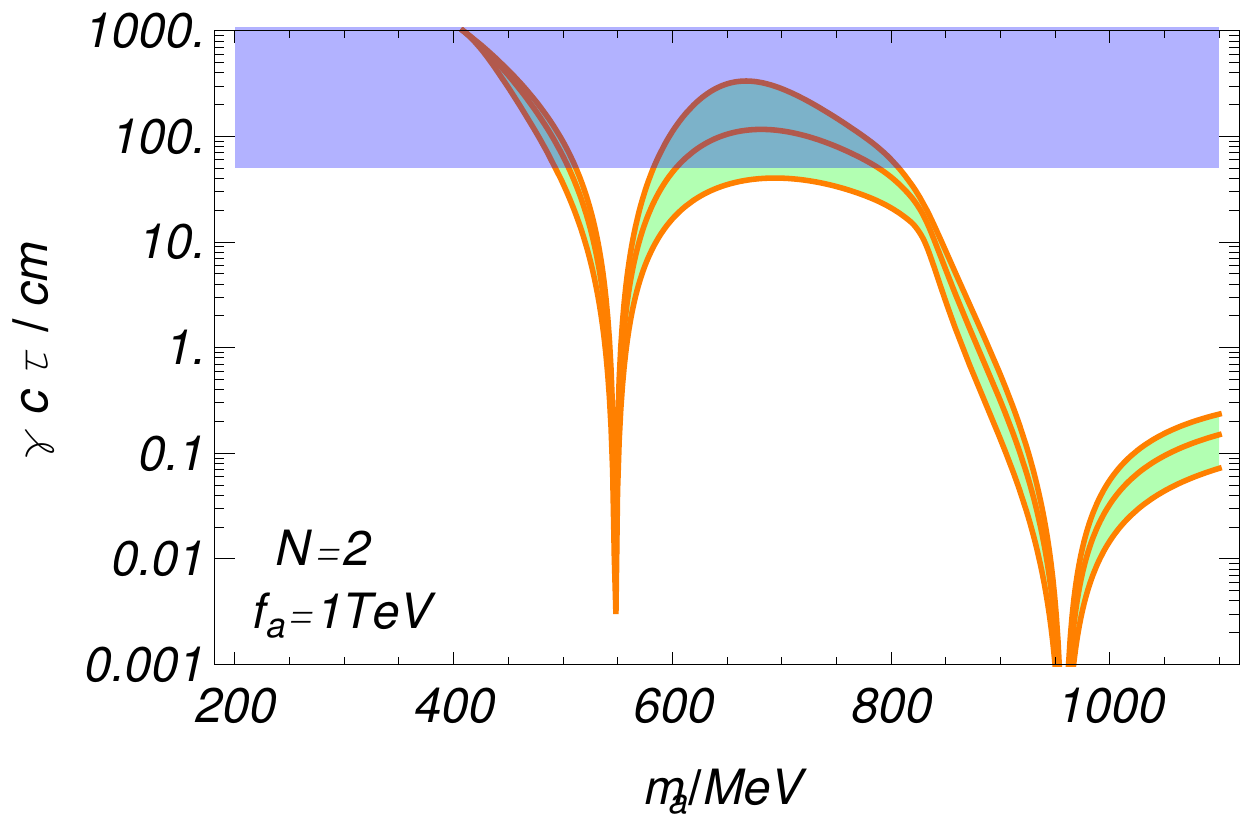}
 \end{center}
\caption{\sl \small
The boosted decay length of an axion for $f_a = 1$\,TeV.
We assume that the axion is produced from a two-body decay 
of a resonance with mass $750$\,GeV and $N$ $2$.  Hence, the 
boost factor $\gamma \simeq 325\,{\rm GeV}/m_a$.
The shaded blue region indicates that the boosted decay width exceeds $50$\,cm.
}
\label{fig:length}
\end{figure}

With these cautions in mind, we simply assume that the acceptance of $a\to 2\gamma$ as
a single photon is of ${\cal O}(1)$, while $a\to 3\pi^0 \to 6\gamma$ is not acceptable.
In Fig.\,\ref{fig:branching}, we show the branching  ratio of the axion into $2\gamma$ for $f_a = 1$\,TeV.
Here, to take account of the ${\cal O}(1)$ uncertainties in the estimation of the widths into $a\to 3\pi$
and $a\to\eta+2\pi$, we adopt the decay widths of those modes scaled from the ones given in Fig.\,\ref{fig:branching} 
by a factor of $1/3$ (top), $1$ (middle), and $3$ (bottom) for each line.
The figure shows that the branching ratio of $a\to 2\gamma$ can be of ${\cal O}(1)$ 
for $400\,{\rm MeV}\lesssim m_a \lesssim 600$\,MeV and $m_a \simeq 800$\,MeV.

It is noted that the decay length of the axion should be sufficiently short to account for the signal.
In Fig.\,\ref{fig:length}, we show the decay length of an axion 
generated from the decay of a $750$\,GeV resonance for $f_a = 1\,$TeV and $N=2$.
The figure shows that the axion decays well before the electromagnetic calorimeter.
Thus, the di-axion signal in which each of the axions decays into a pair of photons can indeed
mimic the diphoton signal.
When the boosted decay length is longer than about $50$\,cm, it means that some of the axions reach 
the electromagnetic calorimeter before they decay.

\begin{figure}[t]
\begin{center}
  \includegraphics[width=0.55\linewidth]{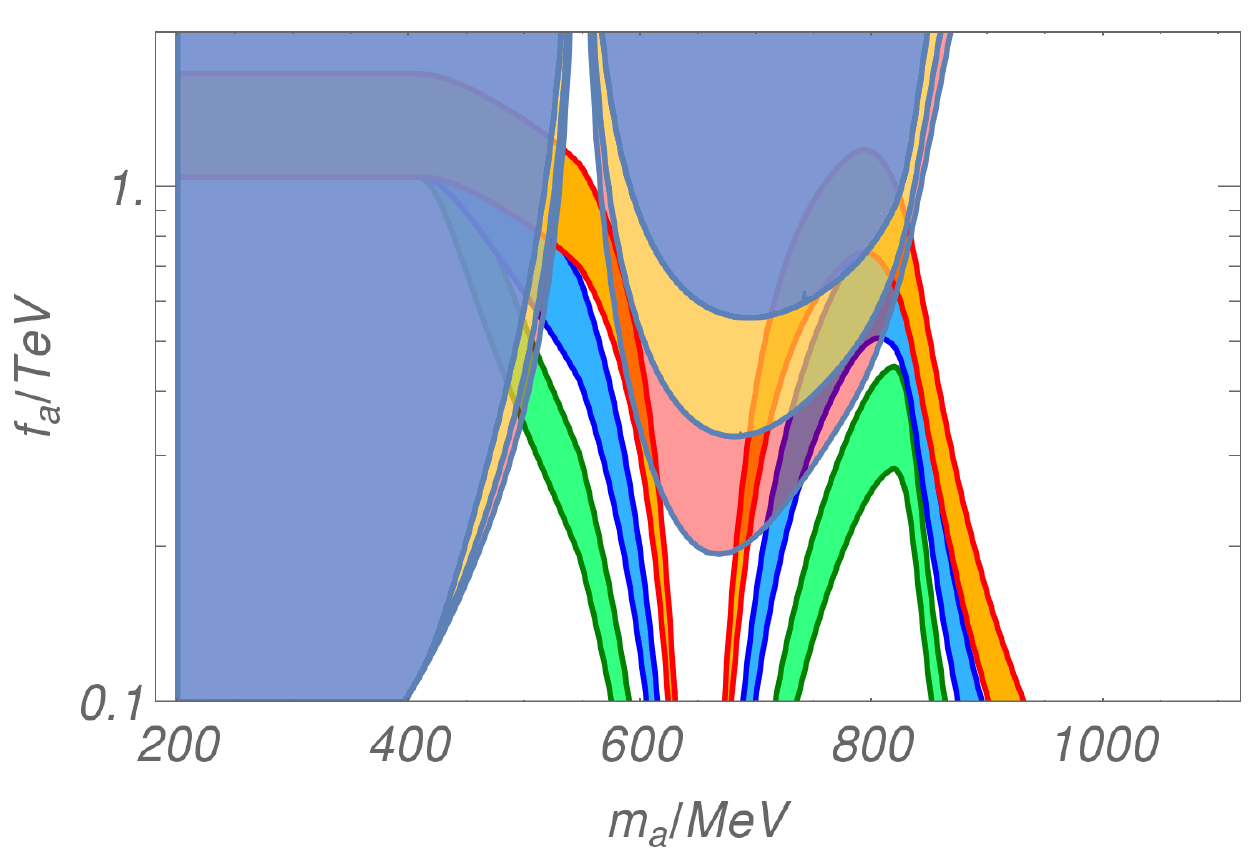}
 \end{center}
\caption{\sl \small
The favored decay constant to reproduce the diphoton excess via the di-axion signal for $N=2$.
Each colored band corresponds to the decay constant that gives the appropriate cross section times
branching ratios in Eq.\,(\ref{eq:signal}).
For each band, we adopt the decay widths of $a\to 3\pi$'s which is 
scaled by a factor of $1/3$ (orange), $1$ (blue) and $3$ (green), respectively.
The shaded regions are excluded from the condition, $c\tau\gamma > 50\,\text{cm}$. The red, yellow and blue regions corresponds where $\Gamma(a\to3\pi)$ is multiplied by $1/3$, $1$ and $3$, respectively.
}
\label{fig:fa}
\end{figure}

By putting all the above discussions together, we now find the preferred range of the decay constant $f_a$ to explain the diphoton excess by the di-axion signal with each of the axions decaying into two photons.  In Fig.\,\ref{fig:fa}, we show the decay constant that satisfies Eq.\,(\ref{eq:signal}) for $N=2$.
To take account of the ${\cal O}(1)$ uncertainties in the estimation of the widths into $a\to 3\pi$ and $a\to \eta + 2 \pi$,
we adopt the decay widths of these modes scaled by a factor of $1/3$ (orange), $1$ (blue) and $3$ (green) 
from the one  given in Fig.\,\ref{fig:branching} for each band.
The figure shows that the signal can be explained for $f_a \simeq 1$\,TeV and $m_a$ around $500\,\text{MeV}$ with $N\gtrsim2$.
It should be noted that the boosted decay length of the axion is longer than $50$\,cm when the axion is too light.

\subsection{Model with multiple extra fermions}
\label{sec:multimatter}
As we have discussed, we may introduce multiple  matters, {\it i.e.} non-zero $N$. One caveat here is the serious domain wall problem, 
as the discrete ${\mathbb Z}_N$ subgroup of the Peccei-Quinn symmetry 
remains unbroken by the anomalies of QCD$^{(}{}'{}^{)}$.%
\footnote{If the signs of Peccei-Quinn charges of the multiple extra fermions are different, it is possible 
to set the domain wall number to be unity even for $N>1$.}
The discrete ${\mathbb Z}_N$ symmetry is eventually broken spontaneously by the VEV of the axion once it attains a potential from QCD$'$ 
effects,
and the domain wall is formed when the spontaneous symmetry breakdown takes place.
If the ${\mathbb Z}_N$ symmetry is an exact symmetry, the domain walls are stable and immediately dominate the energy density of the universe.

As we have discussed at the end of section\,\ref{sec:visible_axion},  the visible heavy axion model
is successful even if the Peccei-Quinn symmetry is explicitly broken by Planck-suppressed operators. 
In the presence of such extra breaking, the discrete $\mathbb Z_N$ symmetry is no longer exact, 
since the explicit breaking generates the differences in the energy densities of vacua at the order of 
\begin{eqnarray}
{\mit \Delta} V \sim \frac{f_a^5}{M_{\rm PL}}\ ,
\end{eqnarray}
and hence the formed domain walls are unstable\,\cite{Vilenkin:1981zs,Preskill:1991kd}.
In fact, the domain walls get accelerated towards the domain with a higher vacuum energy density
\begin{eqnarray}
A \simeq \frac{{\mit \Delta} V}{T_{\rm wall}\,\Lambda'^4}
 \simeq 
\frac{{\mit \Delta} V}{f_a\Lambda'^2} 
 \simeq 
\frac{f_a^4}{M_{\rm PL}\Lambda'^2} \ .
\end{eqnarray}
Here, we approximate the thickness of a domain wall to be $T_{\rm wall} \simeq f_a/\Lambda'^2$.
With such an acceleration much larger than the Hubble parameter at the time of domain wall formation,
the domain walls collapse immediately after their formation.%
\footnote{The domain walls mainly collapse into axions.  The energy density of the emitted axions
are much smaller than that in the thermal bath.}
Therefore, the models with multiple sets of fermions do not cause the domain wall problem in this model, thanks to a small axion decay constant.

Finally, let us emphasize that the extra colored fermions in multiples of three are important to 
evade the dark radiation constraint. 
As we have commented in the previous section, it is safe to consider a model
where the intrinsic axion anomalous couplings to $\gamma$ and $\gamma'$ 
as in Eq.\,(\ref{eq:anoms}) and Eq.\,(\ref{eq:aF'F'}) are vanishing.
For that purpose, we need to introduce extra colorless fermions with the 
opposite Peccei-Quinn charge to cancel the anomalous couplings to  $\gamma$ and $\gamma'$.
The gauge charges of the extra fermions are restricted to allow small mixings to the
Standard Model$^{(}{}'{}^{)}$ fermions to avoid stable colored/charged particles.
As a result, the intrinsic anomalous couplings to  $\gamma$ and $\gamma'$ can be cancelled 
by the colorless contributions only for the extra colored fermions in multiples of three.

\section{Conclusions}
\label{sec:discussions}

In this paper, we have revisited a visible heavy QCD axion model, studied in Ref.~\cite{Fukuda:2015ana}, in light of the recent reports on the $750$\,GeV diphoton resonance by the 
ATLAS and CMS experiments\,\cite{ATLASdiphoton,CMS:2015dxe}. 
In this model, the axion is made heavy with the help of the mirror copied sector of the Standard Model, thereby evading all the astrophysical/cosmological/experimental constraints even for $f_a = {\cal O}(1)$\,TeV while preserving the successful Peccei-Quinn mechanism.
The smallness of the decay constant is highly advantageous when we consider explicit breaking effects
of the Peccei-Quinn symmetry by quantum gravity. 

We identify the $750$\,GeV resonance as the scalar boson associated with spontaneous breakdown of the 
Peccei-Quinn symmetry whose primary decay mode is a pair of the axions.
By carefully examining the decay properties of the axion, we find that its branching ratio into two photons and the boosted decay length can be suitable for explaining the observed diphoton excess 
using the di-axion signal for $f_a \simeq 1$\,TeV.
We also argue that our model allows multiple sets of extra fermions without inducing the domain wall problem.  
Such additional fermions are not only advantageous to explaining the signal but also important to constructing a model
that evades cosmological constraints.

As seen in Fig.\,\ref{fig:length}, the axion mainly decays outside the beam pipe of the LHC experiments.
Since the axion has sizable branching ratios into $3\pi$'s, these modes are expected to leave striking signals inside the detectors, with which we can test the model.

Finally, let us comment on some possibilities to achieve a broad width 
$\Gamma\simeq45\,\text{GeV}$ that is slightly favored by the ATLAS experiment.
As we have discussed in Eq.\,(\ref{eq:dilaton_width}),  such a broad width 
requires a small decay constant $f_a \lesssim 300$\,GeV, which 
predicts too light extra colored particles for $g \lesssim 1$.
To avoid this problem, we need to assume a strong coupling between
$\phi$ and extra fermions, {\it i.e.} $g\gg 1$.
Such a strong coupling blows up immediately above the TeV scale, and hence,
we need some UV completion where $\phi$ and/or $\psi$'s are composite particles.
Such a possibility will be discussed elsewhere.

Another way to achieve a broad width is to introduce extra fermions which 
are not changed under the gauge symmetries of the Standard Model nor 
the ones of the copied sector, so that $\phi$ mainly decays into those 
invisible extra fermions.%
\footnote{Those extra neutral fermion can be a dark matter candidate.}
In this case, we also need to assume a rather large $N$ to make 
the production cross section of $s$ enhanced to explain the diphoton
excess since the branching ratio into $2a$ is small when 
$\phi$ mainly decays into the invisible extra fermions (see {\it e.g.} \cite{Bi:2015uqd,Barducci:2015gtd}
for related discussions).

\section*{Acknowledgements}
HF and MI thank Mihoko Nojiri for useful discussions.
The authors thank Takeo Moroi for pointing out errors in our calculation.
This work is supported in part by the Ministry of Science and Technology of Taiwan under Grant No.~MOST104-2628-M-008-004-MY4 (C.-W.~C); Grants-in-Aid for Scientific Research from the Ministry of Education, Culture, Sports, Science, and Technology (MEXT), Japan, No. 24740151 and No. 25105011 (M.~I.) as well as No. 26104009 (T.~T.~Y.); 
Grant-in-Aid No. 26287039 (M.~I. and T.~T.~Y.) from the Japan Society for the Promotion of Science (JSPS); and by the World Premier International Research Center Initiative (WPI), MEXT, Japan (M.~I., and T.~T.~Y.).
The work of H.F. is supported in part by a Research Fellowship for Young Scientists from
the Japan Society for the Promotion of Science (JSPS).
MI is grateful to the Center for Mathematics and Theoretical Physics and Department of Physics 
of National Central University for hospitality where this work is initiated.

\appendix
\section{Axion Decay Width}
\label{sec:chpt}
In this appendix, we summarize the decay width formulas for the axion.
\subsection{$a\to \gamma\gamma$}
According to Eq.\,(\ref{eq:combine}), the axion coupling to two photon is given by
\begin{eqnarray}
{\cal L} = {C_{a\gamma\gamma}}\frac{a}{f_a} \frac{\alpha_{\rm QED}}{8\pi} F\tilde{F}\ ,
\end{eqnarray}
with an effective coefficient
\begin{eqnarray}
C_{a\gamma\gamma} = \frac{8}{3} + \frac{f_a}{f_\pi}{\varepsilon}_{a\eta'} C_{\eta' \gamma\gamma} 
+ \frac{f_a}{f_\pi} {\varepsilon}_{a\eta} C_{\eta \gamma\gamma} \ .
\end{eqnarray}
Here, we define
\begin{eqnarray}
C_{\eta\gamma\gamma} &=&\frac{2}{\sqrt3}\left(\frac{f_\pi}{f_{\eta^0}}\cos\theta -\sqrt{8}\frac{f_\pi}{f_{\pi_9}}\sin\theta\right)\ , \\
C_{\eta'\gamma\gamma} &=& \frac{2}{\sqrt3}\left(\frac{f_\pi}{f_{\eta^0}}\sin\theta +\sqrt{8}\frac{f_\pi}{f_{\pi_9}}\cos\theta\right)\ .
\end{eqnarray}
With the effective coefficient, the axion decay width is given by
\begin{eqnarray}
\Gamma = \frac{1}{4\pi}\left(\frac{\alpha_{\rm QED}}{8\pi}\right)^2 C_{a\gamma\gamma}^2 \frac{m_a^3}{f_a^2}\ .
\end{eqnarray}

\subsection{$a\to \rho+\gamma$}
To estimate the $a\to\rho+\gamma$ decay, we parameterize the $\rho+\gamma$ couplings to $\eta$ and $\eta'$ by effective interactions (see, {\it e.g.}, Ref.~\cite{Ball:1995zv}),
\begin{eqnarray}
{\cal L}= \frac{1}{2}\frac{\eta_8}{f}  c_{\eta_8\rho}F_\rho\tilde{F} +
\frac{1}{2}\frac{\eta_0}{f} c_{\eta_0\rho}F_\rho\tilde{F} + (F_\rho \leftrightarrow F)\ .
\end{eqnarray}
From the observed decay widths\,\cite{Agashe:2014kda}
\begin{eqnarray}
\Gamma_{\rho\to\eta+\gamma} \simeq  44\,{\rm keV}\ , \quad
\Gamma_{\eta' \to\rho+\gamma} =  57\,{\rm keV}\ ,
\end{eqnarray}
we obtain
\begin{eqnarray}
c_{\eta_8\rho} &=& 9.2 \times 10^{-3}\ , \nonumber\\
c_{\eta_0\rho} &=& 1.3 \times 10^{-2}\ .
\end{eqnarray}
By suitably combining the coefficients, we derive
\begin{eqnarray}
{\cal L}= \frac{1}{2}\frac{a^0}{f_a} c_{a\rho}
F_\rho\tilde{F}  + (F_\rho \leftrightarrow F)\ ,
\end{eqnarray}
with
\begin{eqnarray}
c_{a\rho} &\simeq& {\varepsilon}_{a\eta} \left(\frac{f_a}{f_{8}}c_{\eta_8\rho}\cos\theta -\frac{f_a}{f_{0}}c_{\eta_0\rho}\sin\theta\right) +
 {\varepsilon}_{a\eta'}  \left(\frac{f_a}{f_{8}}c_{\eta_8\rho}\sin\theta +\frac{f_a}{f_{0}}c_{\eta_0\rho}\cos\theta\right)\ .
\end{eqnarray}
With this coefficient, we have 
\begin{eqnarray}
\Gamma = \frac{1}{2\pi}\frac{c_{a\rho}^2}{f_a^2}\frac{(m_{a}^2-M_\rho^2 )^3}{m_{a}^3}\ .
\end{eqnarray}

\subsection{$a\to3\pi$, $a\to \eta+2\pi$}
Since the decay widths of $\eta$ and $\eta'$ are not well reproduced by the chiral perturbation theory, 
we estimate the axion decay width by using an approximate amplitude of $\eta $ and $\eta'$ into $3\pi$, 
\begin{eqnarray}
|{\cal A}(a\to 3\pi)| &=& | {\varepsilon}_{a\eta} \bar{\cal A}(\eta\to 3 \pi^0)|  +  | {\varepsilon}_{a\eta'} \bar{\cal A}(\eta'\to 3 \pi^0)|  \ , \\
|{\cal A}(a\to \eta+2\pi)| &=& | {\varepsilon}_{a\eta'} \bar{\cal A}(\eta'\to \eta+2\pi^0)|  \ . 
\end{eqnarray}
Here, $| \bar{\cal A}|$'s on the right hand side are estimated by dividing the decay widths \cite{Agashe:2014kda} by the corresponding phase space volumes and taking the square root\,\footnote{The amplitudes come from scalar $4$\mbox{-}point interactions and are dimensionless.}:
\begin{eqnarray}
\label{eq:3pi}
\left|\bar{\cal A}(\eta \to \pi^0\pi^+\pi^-)\right| &\simeq& 0.26\ , \quad
\left|\bar{\cal A}(\eta \to 3\pi^0)\right| \simeq  0.30\ ,\\
\left|\bar{\cal A}(\eta' \to \pi^0\pi^+\pi^-)\right|  &\simeq& 0.15\ , \quad
\left|\bar{\cal A}(\eta' \to 3\pi^0)\right|  \simeq 0.11 \ ,\\
\left|\bar{\cal A}(\eta' \to \eta\pi^+ \pi^-)\right| &\simeq& 6.7 \ ,  \, \quad
\left|\bar{\cal A}(\eta' \to \eta\pi^0 \pi^0)\right|\simeq 4.5\ .
\label{eq:3pi2}
\end{eqnarray}

Since we do not know how the $\eta$ and $\eta'$ modes interfere with each other, we simply assume 
constructive interference.
Accordingly, our estimations of the decay widths for $a\to 3\pi$ and $a\to \eta + 2\pi$ have ${\cal O}(1)$ uncertainties.


\begin{thebibliography}{99}
	
\bibitem{Kobayashi:1973fv} 
M.~Kobayashi and T.~Maskawa,
Prog.\ Theor.\ Phys.\  {\bf 49}, 652 (1973).
doi:10.1143/PTP.49.652

\bibitem{Sakharov:1967dj} 
A.~D.~Sakharov,
Pisma Zh.\ Eksp.\ Teor.\ Fiz.\  {\bf 5}, 32 (1967)
[JETP Lett.\  {\bf 5}, 24 (1967)]
[Sov.\ Phys.\ Usp.\  {\bf 34}, 392 (1991)]
[Usp.\ Fiz.\ Nauk {\bf 161}, 61 (1991)].
doi:10.1070/PU1991v034n05ABEH002497

\bibitem{Fukugita:1986hr} 
M.~Fukugita and T.~Yanagida,
Phys.\ Lett.\ B {\bf 174}, 45 (1986).
doi:10.1016/0370-2693(86)91126-3

\bibitem{Peccei:1977hh} 
	R.~D.~Peccei and H.~R.~Quinn,
	Phys.\ Rev.\ Lett.\  {\bf 38}, 1440 (1977).
	doi:10.1103/PhysRevLett.38.1440

\bibitem{Peccei:1977ur} 
R.~D.~Peccei and H.~R.~Quinn,
Phys.\ Rev.\ D {\bf 16}, 1791 (1977).
doi:10.1103/PhysRevD.16.1791

\bibitem{Weinberg:1977ma} 
S.~Weinberg,
Phys.\ Rev.\ Lett.\  {\bf 40}, 223 (1978).
doi:10.1103/PhysRevLett.40.223




\bibitem{Wilczek:1977pj} 
F.~Wilczek,
Phys.\ Rev.\ Lett.\  {\bf 40}, 279 (1978).
doi:10.1103/PhysRevLett.40.279

\bibitem{Agashe:2014kda} 
  K.~A.~Olive {\it et al.} [Particle Data Group Collaboration],
  Chin.\ Phys.\ C {\bf 38}, 090001 (2014).
  doi:10.1088/1674-1137/38/9/090001
  
\bibitem{Kim:1979if} 
J.~E.~Kim,
``Weak Interaction Singlet and Strong CP Invariance,''
Phys.\ Rev.\ Lett.\  {\bf 43}, 103 (1979).
doi:10.1103/PhysRevLett.43.103

\bibitem{Shifman:1979if} 
M.~A.~Shifman, A.~I.~Vainshtein and V.~I.~Zakharov,
``Can Confinement Ensure Natural CP Invariance of Strong Interactions?,''
Nucl.\ Phys.\ B {\bf 166}, 493 (1980).
doi:10.1016/0550-3213(80)90209-6

\bibitem{Zhitnitsky:1980tq} 
A.~R.~Zhitnitsky,
``On Possible Suppression of the Axion Hadron Interactions. (In Russian),''
Sov.\ J.\ Nucl.\ Phys.\  {\bf 31}, 260 (1980)
[Yad.\ Fiz.\  {\bf 31}, 497 (1980)].

\bibitem{Dine:1981rt} 
M.~Dine, W.~Fischler and M.~Srednicki,
``A Simple Solution to the Strong CP Problem with a Harmless Axion,''
Phys.\ Lett.\ B {\bf 104}, 199 (1981).
doi:10.1016/0370-2693(81)90590-6

\bibitem{Tye:1981zy} 
For instance, see
S.~H.~H.~Tye,
Phys.\ Rev.\ Lett.\  {\bf 47}, 1035 (1981).
doi:10.1103/PhysRevLett.47.1035

\bibitem{Rubakov:1997vp} 
V.~A.~Rubakov,
JETP Lett.\  {\bf 65}, 621 (1997)
doi:10.1134/1.567390
[hep-ph/9703409].

\bibitem{Fukuda:2015ana} 
H.~Fukuda, K.~Harigaya, M.~Ibe and T.~T.~Yanagida,
Phys.\ Rev.\ D {\bf 92}, no. 1, 015021 (2015)
doi:10.1103/PhysRevD.92.015021
[arXiv:1504.06084 [hep-ph]].

\bibitem{Berezhiani:2000gh} 
  Z.~Berezhiani, L.~Gianfagna and M.~Giannotti,
  Phys.\ Lett.\ B {\bf 500}, 286 (2001)
  doi:10.1016/S0370-2693(00)01392-7
  [hep-ph/0009290].
\bibitem{Hook:2014cda} 
  A.~Hook,
  Phys.\ Rev.\ Lett.\  {\bf 114}, no. 14, 141801 (2015)
  doi:10.1103/PhysRevLett.114.141801
  [arXiv:1411.3325 [hep-ph]].

\bibitem{Albaid:2015axa} 
  A.~Albaid, M.~Dine and P.~Draper,
  JHEP {\bf 1512}, 046 (2015)
  doi:10.1007/JHEP12(2015)046
  [arXiv:1510.03392 [hep-ph]].



\bibitem{ATLASdiphoton} 
The ATLAS collaboration,
ATLAS-CONF-2015-081.

\bibitem{CMS:2015dxe} 
CMS Collaboration [CMS Collaboration],
collisions at 13TeV,''
CMS-PAS-EXO-15-004.

\bibitem{Higaki:2015jag} 
T.~Higaki, K.~S.~Jeong, N.~Kitajima and F.~Takahashi,
Phys.\ Lett.\ B {\bf 755}, 13 (2016)
doi:10.1016/j.physletb.2016.01.055
[arXiv:1512.05295 [hep-ph]].
\bibitem{Kim:2015xyn} 
  J.~E.~Kim,
  Phys.\ Lett.\ B {\bf 755}, 190 (2016)
  doi:10.1016/j.physletb.2016.02.016
  [arXiv:1512.08467 [hep-ph]].
  
\bibitem{Aparicio:2016iwr} 
L.~Aparicio, A.~Azatov, E.~Hardy and A.~Romanino,
arXiv:1602.00949 [hep-ph].

\bibitem{Ellwanger:2016qax} 
U.~Ellwanger and C.~Hugonie,
arXiv:1602.03344 [hep-ph].

  
\bibitem{Knapen:2015dap} 
S.~Knapen, T.~Melia, M.~Papucci and K.~Zurek,
arXiv:1512.04928 [hep-ph].

\bibitem{Agrawal:2015dbf} 
P.~Agrawal, J.~Fan, B.~Heidenreich, M.~Reece and M.~Strassler,
arXiv:1512.05775 [hep-ph].

\bibitem{Chala:2015cev} 
M.~Chala, M.~Duerr, F.~Kahlhoefer and K.~Schmidt-Hoberg,
Phys.\ Lett.\ B {\bf 755}, 145 (2016)
doi:10.1016/j.physletb.2016.02.006
[arXiv:1512.06833 [hep-ph]].

\bibitem{Dasgupta:2016wxw} 
B.~Dasgupta, J.~Kopp and P.~Schwaller,
arXiv:1602.04692 [hep-ph].


\bibitem{Seesaw}
T.~Yanagida, in Proceedings of the Workshop on Unified Theories and Baryon Number in the Universe,
eds.\ O.~Sawada and A.~Sugamoto (KEK report 79-18, 1979);
M.~Gell-Mann, P.~Ramond and R.~Slansky, 
in Sanibel Symposium, Palm Coast, Fla., Feb 1979.
Also see,
P.~Minkowski,
Phys.\ Lett.\  B {\bf 67}, 421 (1977).

\bibitem{Bergsma:1985qz} 
F.~Bergsma {\it et al.} [CHARM Collaboration],
Phys.\ Lett.\ B {\bf 157}, 458 (1985).
doi:10.1016/0370-2693(85)90400-9

\bibitem{Aad:2014efa} 
G.~Aad {\it et al.} [ATLAS Collaboration],
JHEP {\bf 1411}, 104 (2014)
doi:10.1007/JHEP11(2014)104
[arXiv:1409.5500 [hep-ex]].

\bibitem{Aad:2015tba} 
G.~Aad {\it et al.} [ATLAS Collaboration],
Phys.\ Rev.\ D {\bf 92}, no. 11, 112007 (2015)
doi:10.1103/PhysRevD.92.112007
[arXiv:1509.04261 [hep-ex]].

\bibitem{Khachatryan:2015gza} 
V.~Khachatryan {\it et al.} [CMS Collaboration],
arXiv:1507.07129 [hep-ex].

\bibitem{Khachatryan:2015oba} 
V.~Khachatryan {\it et al.} [CMS Collaboration],
Phys.\ Rev.\ D {\bf 93}, no. 1, 012003 (2016)
doi:10.1103/PhysRevD.93.012003
[arXiv:1509.04177 [hep-ex]].


\bibitem{Draper:2012xt} 
P.~Draper and D.~McKeen,
Phys.\ Rev.\ D {\bf 85}, 115023 (2012)
doi:10.1103/PhysRevD.85.115023
[arXiv:1204.1061 [hep-ph]].

\bibitem{Ellis:2012zp} 
S.~D.~Ellis, T.~S.~Roy and J.~Scholtz,
Phys.\ Rev.\ D {\bf 87}, no. 1, 014015 (2013)
doi:10.1103/PhysRevD.87.014015
[arXiv:1210.3657 [hep-ph]].

\bibitem{Ellis:2012sd} 
S.~D.~Ellis, T.~S.~Roy and J.~Scholtz,
Phys.\ Rev.\ Lett.\  {\bf 110}, no. 12, 122003 (2013)
doi:10.1103/PhysRevLett.110.122003
[arXiv:1210.1855 [hep-ph]].

\bibitem{Dobrescu:2000jt} 
B.~A.~Dobrescu, G.~L.~Landsberg and K.~T.~Matchev,
Phys.\ Rev.\ D {\bf 63}, 075003 (2001)
doi:10.1103/PhysRevD.63.075003
[hep-ph/0005308].

\bibitem{Toro:2012sv} 
N.~Toro and I.~Yavin,
Phys.\ Rev.\ D {\bf 86}, 055005 (2012)
doi:10.1103/PhysRevD.86.055005
[arXiv:1202.6377 [hep-ph]].

\bibitem{Chang:2006bw} 
S.~Chang, P.~J.~Fox and N.~Weiner,
Phys.\ Rev.\ Lett.\  {\bf 98}, 111802 (2007)
doi:10.1103/PhysRevLett.98.111802
[hep-ph/0608310].

\bibitem{Curtin:2013fra} 
D.~Curtin {\it et al.},
Phys.\ Rev.\ D {\bf 90}, no. 7, 075004 (2014)
doi:10.1103/PhysRevD.90.075004
[arXiv:1312.4992 [hep-ph]].



\bibitem{Martin:2009iq} 
  A.~D.~Martin, W.~J.~Stirling, R.~S.~Thorne and G.~Watt,
  Eur.\ Phys.\ J.\ C {\bf 63}, 189 (2009)
  [arXiv:0901.0002 [hep-ph]].
  
\bibitem{Baglio:2010ae} 
  J.~Baglio and A.~Djouadi,
  JHEP {\bf 1103}, 055 (2011)
  doi:10.1007/JHEP03(2011)055
  [arXiv:1012.0530 [hep-ph]].

  
\bibitem{Beisert:2003zs} 
  N.~Beisert and B.~Borasoy,
  Nucl.\ Phys.\ A {\bf 716}, 186 (2003)
  doi:10.1016/S0375-9474(02)01585-3
  [hep-ph/0301058].

\bibitem{Donoghue:1986wv} 
  J.~F.~Donoghue, B.~R.~Holstein and Y.~C.~R.~Lin,
  Phys.\ Rev.\ Lett.\  {\bf 55}, 2766 (1985)
  [Phys.\ Rev.\ Lett.\  {\bf 61}, 1527 (1988)].
  doi:10.1103/PhysRevLett.55.2766



\bibitem{Bijnens:1988kx} 
  J.~Bijnens, A.~Bramon and F.~Cornet,
  Phys.\ Rev.\ Lett.\  {\bf 61}, 1453 (1988).
  doi:10.1103/PhysRevLett.61.1453
\bibitem{Ecker:2015uoa} 
  G.~Ecker,
  arXiv:1510.01634 [hep-ph].
  
  
\bibitem{Ablikim:2015cmz} 
M.~Ablikim {\it et al.} [BESIII Collaboration],
Phys.\ Rev.\ D {\bf 92}, 012014 (2015)
doi:10.1103/PhysRevD.92.012014
[arXiv:1506.05360 [hep-ex]].

\bibitem{Ade:2015xua} 
  P.~A.~R.~Ade {\it et al.} [Planck Collaboration],
  arXiv:1502.01589 [astro-ph.CO].

\bibitem{ATLASNOTE}
  The ATLAS Collaboration, ATL-PHYS-PUB-2009-006
\bibitem{withMihoko}
  H.~Fukuda, M.~Ibe and M.~Noriji in preparation.
  
\bibitem{Aad:2008zzm} 
G.~Aad {\it et al.} [ATLAS Collaboration],
JINST {\bf 3}, S08003 (2008).
doi:10.1088/1748-0221/3/08/S08003

\bibitem{Chatrchyan:2008aa} 
S.~Chatrchyan {\it et al.} [CMS Collaboration],
JINST {\bf 3}, S08004 (2008).
doi:10.1088/1748-0221/3/08/S08004

\bibitem{Vilenkin:1981zs} 
  A.~Vilenkin,
  Phys.\ Rev.\ D {\bf 23}, 852 (1981).
  doi:10.1103/PhysRevD.23.852
\bibitem{Preskill:1991kd} 
  J.~Preskill, S.~P.~Trivedi, F.~Wilczek and M.~B.~Wise,
  Nucl.\ Phys.\ B {\bf 363}, 207 (1991).
  doi:10.1016/0550-3213(91)90241-O
\bibitem{Bi:2015uqd} 
  X.~J.~Bi, Q.~F.~Xiang, P.~F.~Yin and Z.~H.~Yu,
  arXiv:1512.06787 [hep-ph].
\bibitem{Barducci:2015gtd} 
  D.~Barducci, A.~Goudelis, S.~Kulkarni and D.~Sengupta,
  arXiv:1512.06842 [hep-ph].
  
\bibitem{Ball:1995zv} 
  P.~Ball, J.~M.~Frere and M.~Tytgat,
  Phys.\ Lett.\ B {\bf 365}, 367 (1996)
  doi:10.1016/0370-2693(95)01287-7
  [hep-ph/9508359].

\end{thebibliography}
\end{document}